\documentclass[%
 aip,
 rsi,
amsmath,amssymb,
reprint,
floatfix
]{revtex4-1}

\usepackage{graphicx}
\usepackage{dcolumn}
\usepackage{bm}

\usepackage[utf8]{inputenc}
\usepackage{mathptmx}
\usepackage{multirow}
\usepackage{braket}

\usepackage[english]{babel}
\usepackage{xcolor}
\colorlet{RED}{red}
\colorlet{BLUE}{blue}
\usepackage{dcolumn}
\usepackage{bm}
\usepackage[version=4]{mhchem} 
\usepackage{acronym}
\usepackage{adjustbox}
\usepackage{float}
\usepackage{tikz}
\usepackage{microtype} 
\usepackage{algorithm}
\usepackage{algpseudocode}

\usetikzlibrary{calc,shapes.geometric,decorations.pathmorphing,patterns}

\definecolor{background-color}{gray}{0.98}
\usepackage[margin=2.3cm,bmargin=1cm,footnotesep=1cm]{geometry}

\begin{document}

\title{Impact of high-rank excitations on accuracy of the unitary coupled cluster downfolding formalism}

\author{Karol Kowalski}
\email{karol.kowalski@pnnl.gov}
\affiliation{Physical Sciences Division, Pacific Northwest National Laboratory, Richland, Washington 99354, United States}

\author{Bo Peng}
\affiliation{Physical Sciences Division, Pacific Northwest National Laboratory, Richland, Washington 99354, United States}

 \author{Nicholas P. Bauman}
\affiliation{Physical Sciences Division, Pacific Northwest National Laboratory, Richland, Washington 99354, United States}

\begin{abstract}
    In this paper, we evaluate the accuracy of the Hermitian form of the downfolding procedure utilizing the double unitary coupled cluster Ansatz (DUCC) on the H6 and H8 benchmark systems. The computational infrastructure employs the occupation-number-representation codes to construct the matrix representation of arbitrary second-quantized operators, enabling the exact representation of exponentials of various operators. The tests utilize external excitations estimated from standard single-reference coupled cluster methods (SR-CC) to demonstrate that higher-rank SR-CC external amplitudes were necessary to describe the energies in the strongly correlated regime adequately. We show that this approach can offset problems of the corresponding SR-CC theories associated with losing the variational character of corresponding energies. 
\end{abstract}

\maketitle

\section{Introduction}
Applying many-body methods for dimensionality/cost reduction (DCR) of {\it ab-initio} formulations is imperative in expanding the range of system sizes amenable to accurate many-body formulations in chemistry and material sciences. Additionally, these techniques are vital in effectively using early quantum computing resources, commonly referred to as the noisy intermediate-scale quantum devices (NISQ). 
\cite{Tyler2020,csahinouglu2021hamiltonian,EffectiveH2022,huang2023leveraging}
DCR methods primarily focus on minimizing the number of qubits required to represent a given quantum problem. 
One should mention several techniques developed to take full advantage of the ubiquitous Variational Quantum Eigensolvers (VQE) approach 
\cite{peruzzo2014variational,mcclean2016theory,romero2018strategies,Kandala2017,kandala2019,izmaylov2019unitary,lang2020unitary,grimsley2019adaptive,grimsley2019trotterized,mcardle2020quantum,Love2021,tilly2022variational}
in addressing problems beyond the situation where few electrons are correlated.

In the context of quantum algorithms for quantum chemistry, the utilization of DCR techniques  is linked to the partitioning of electron correlation effects into static and dynamic partitions. In terms of methodology, the coupled cluster (CC) formalism 
\cite{coester58_421,coester60_477,cizek66_4256,shavitt72,mukherjee1975correlation,adams79,purvis82_1910,jezmonk,mrcclyakh}
offers an effective means of describing these effects in a many-body language. Although static effects can be incorporated for small-scale systems using presently accessible quantum hardware, the implementation of necessary dynamical correlation effects, which typically involve numerous parameters with minute values, remains beyond the scope of contemporary quantum computing technologies.

We recently introduced and tested downfolding techniques based on the double unitary coupled cluster Ansatz (DUCC) \cite{bauman2019downfolding} to address the abovementioned problem. The downfolding procedure utilizes the properties of the ground-state  DUCC Ansatz, which in analogy to single-reference sub-system embedding sub-algebras (SES-CC),
\cite{safkk,kowalski2021dimensionality} allows to construct of effective Hamiltonians that integrate out-of-active-space degrees of freedom usually identified with dynamical amplitudes. In contrast to the SES-CC approach, the DUCC formalism yields the Hermitian form of the effective Hamiltonian in the active space. 

The DUCC-driven downfolded Hamiltonians are critical components of hybrid computing. Classical computing resources are employed to calculate the second quantized form of effective Hamiltonians, and quantum computing is invoked to diagonalize them in active spaces that ideally match the available quantum resources. This type of approach provides much needed algorithmic transition mechanism from current NISQ technologies 
\cite{preskill2018quantum}
to mature error-corrected quantum computers of the future, where the size of the active space is adjusted to the available quantum resources. For this purpose, several approximations were tested to validate the efficiency of the downfolding procedure. These approximations, due to the non-commutativity of the components defining DUCC cluster operators,  were based on the finite low-rank commutator expansions, the limited rank of interactions included in the downfolded Hamiltonians
(one- and two-body interactions), and simple form of the external amplitudes extracted from the single-reference CC (SR-CC) model with singles and doubles (CCSD).\cite{purvis82_1910} 

Our team has recently developed a novel full configuration interaction (FCI) code called stringMB, which employs a string-based approach to emulate quantum systems and represent operators in matrix form. This code has been integrated into the NWChem software, enabling us to (1) work with the exact  representations of operator exponents and (2) leverage various sources for external CC amplitudes. Consequently, we have a unique opportunity to study the exact nature of downfolded Hamiltonians using the DUCC method. In this study, we investigate the impact of higher-rank external excitations obtained through CCSD,\cite{purvis82_1910} CCSDT,\cite{scuseria_ccsdt,ccsdt_noga,ccsdt_noga_err} and CCSDTQ \cite{ccsdtq_nevin,doi:10.1063/1.467143,Kucharski1991} simulations, as well as the active space size, on the accuracy of ground-state energies for small benchmark systems H6 and H8 representing linear chains of hydrogen atoms.

\section{Theory}
The DUCC formulations have been amply discussed in recent papers (see Refs.\cite{bauman2019downfolding,downfolding2020t,bauman2022coupled}). Here we overview only the salient features of these approaches. While the SES-CC technique \cite{safkk,kowalski2021dimensionality} forms the basis for non-Hermitian downfolding, the DUCC expansions provide its Hermitian formulations. 
The Hermitian form of the downfolded Hamiltonian is obtained as a consequence of utilizing active-space-dependent DUCC representation of the wave function
\begin{equation}
        |\Psi\rangle=e^{\sigma_{\rm ext}} e^{\sigma_{\rm int}}|\Phi\rangle \;,
\label{ducc1}
\end{equation}
where $\sigma_{\rm ext}$ and $\sigma_{\rm int}$, referred to as the external and internal cluster operators,  are general-type anti-Hermitian operators
\begin{eqnarray}
\sigma_{\rm int}^{\dagger} &=&  -\sigma_{\rm int} \;, \label{sintah} \\
\sigma_{\rm ext}^{\dagger} &=&  -\sigma_{\rm ext} \;. \label{sintah2}
\end{eqnarray} 
In analogy to the non-Hermitian case, the $\sigma_{\rm ext}$ and $\sigma_{\rm int}$ operators are defined by parameters carrying active spin-orbital labels only and at least one in-active spin-orbital label, respectively. 
The DUCC Ansatz falls into a broad class of active space coupled cluster methods.
\cite{oliphant1991multireference,nevin1,pnl93}

The use of the  DUCC Ansatz (\ref{ducc1}), in analogy to the SES-CC case, leads to an alternative way of determining energy, which can be obtained by solving  active-space Hermitian eigenvalue problem:
\begin{equation}
        H^{\rm eff} e^{\sigma_{\rm int}} |\Phi\rangle = E e^{\sigma_{\rm int}}|\Phi\rangle,
\label{duccstep2}
\end{equation}
where
\begin{equation}
        H^{\rm eff} = (P+Q_{\rm int}) \bar{H}_{\rm ext} (P+Q_{\rm int})
\label{equivducc}
\end{equation}
and 
\begin{equation}
        \bar{H}_{\rm ext} =e^{-\sigma_{\rm ext}}H e^{\sigma_{\rm ext}}.
\label{duccexth}
\end{equation}
When the external cluster amplitudes are known (or can be effectively approximated), the energy (or its approximation) can be calculated by diagonalizing the Hermitian effective/downfolded Hamiltonian (\ref{equivducc}) in the active space using various quantum or classical diagonalizers. 
The $Q_{\rm int}$ operator is a projection onto excited (with respect to $|\Phi\rangle$) excited configurations in complete active space (CAS) and the projection onto the reference function is denoted as $P$.

For quantum computing applications second-quantized representation of $H^{\rm eff}$ is required. 
In the light of the non-commuting character of components defining $\sigma_{\rm ext}$ operator, to this end, one has to rely on the finite-rank commutator expansions, i.e., 
\begin{equation}
\bar{H}_{\rm ext}
\simeq
H +  \sum_{i=1}^{\rm Max_R}  \frac{1}{i!}[
   \ldots [H,\sigma_{\rm ext}],\ldots ],\sigma_{\rm ext}]_i\;,
   \label{comm1}
\end{equation}
where ${\rm Max_R}$ stands for the length of commutator expansion.
Due to the numerical costs associated with the contractions of multi-dimensional tensors, only approximations based on including low-rank commutators are feasible. In recent studies, approximations based on single, double, and part of triple commutators were explored where one- and two-body interactions were retained in the second quantized form of $H^{\rm eff}$.
In practical applications, one also has to determine the approximate form of $\sigma_{\rm ext}$. For practical reasons we used the following approximation
\begin{equation}
    \sigma_{\rm ext} \simeq T_{\rm ext} - T_{\rm ext}^{\dagger} \;,
    \label{sigext}
\end{equation}
where $T_{\rm ext}$ can be defined through the external parts of the typical SR-CC cluster operators. 

Given the progress achieved in the development of the Hermitian form of the downfolded Hamiltonians, addressing two pressing questions  (1) what is the  impact of the choice of the $T_{\rm ext}$ on the quality of ground-state energy of $H^{\rm eff}$? and 
(2) what are the energy values corresponding to the untruncated (exact) form of the $H^{\rm eff}$? play a pivotal role in further understanding and advances of CC downfolding techniques. We answer these questions using stringMB code that allows us to deal with the exact matrix representations of second quantized operators and their functions in the FCI space.

\section{Implementation}
For interacting fermionic systems,the action of the creation/annihilation 
operators for the electron in $p$-th spin-orbital ($a_p/a_p^{\dagger}$) on the Slater determinants can be conveniently described using  occupation number representation, where each Slater determinant is represented as a vector
\begin{equation}
|n_M \; n_{M-1}\;  \ldots\;  n_{i+1}\; n_i \;n_{i-1} \;\ldots \;n_1 \rangle
\label{onr}
\end{equation} 
where occupation numbers $n_i$ are equal to either 1 (electron occupies $i$-th spin orbital) or 0 (no electron is occupying $i$-th spin orbital). In (\ref{onr}), $M$ stands for the total number of spin-orbitals used to describe quantum system and $M=2n$, where $n$ is the number of orbitals.  

The following formulas give the non-trivial action of creation/annihilation operators on the state vectors
\begin{widetext}
\begin{eqnarray}
a_i^{\dagger} |n_M \; n_{M-1} \; \ldots \;n_{i+1} \; 0 \; n_{i-1}\; \ldots n_1\; \rangle  &=& (-1)^{\sum_{k=1}^{i-1} n_k} 
|n_M \; n_{M-1} \; \ldots \;n_{i+1} \; 1 \; n_{i-1}\; \ldots n_1\; \rangle
\label{onr1} \;\;\;\;\;\; \\
a_i |n_M \; n_{M-1} \; \ldots \;n_{i+1} \; 1 \; n_{i-1}\; \ldots n_1\; \rangle  &=& (-1)^{\sum_{k=1}^{i-1} n_k} 
|n_M \; n_{M-1} \; \ldots \;n_{i+1} \; 0 \; n_{i-1}\; \ldots n_1\; \rangle .
\label{onr2}
\end{eqnarray} 
\end{widetext}
Using the occupation-number representation, the stringMB code  allows one to construct a matrix representation (${\bf A}$) 
of general second-quantized operators $A$, where $A$ can be identified with electronic Hamiltonian, the external part of the cluster operator $T_{\rm ext}$, 
and exponents of $T_{\rm ext}-
T_{\rm ext}^{\dagger}$, i.e., 
\begin{eqnarray}
H &\rightarrow & {\bf H} \;\;, \label{hmat} \\
T_{\rm ext} &\rightarrow & {\bf T}_{\rm ext} \;\;, \label{textmat} \\
e^{\sigma_{\rm ext}} \simeq
e^{T_{\rm ext}-T_{\rm ext}^{\dagger}}
&\rightarrow &  e^{{\bf T}_{\rm ext}-{\bf T}_{\rm ext}^{\dagger}},  \label{expsigma1} \\
e^{-\sigma_{\rm ext}} \simeq
e^{-(T_{\rm ext}-T_{\rm ext}^{\dagger})}
&\rightarrow &  e^{-({\bf T}_{\rm ext}-{\bf T}_{\rm ext}^{\dagger})},  \label{expsigma2} \\
\bar{H}_{\rm ext} &\rightarrow & \bar{\bf H}_{\rm ext}, \label{hextma} \\
 H^{\rm eff}  &\rightarrow & {\bf H}^{\rm eff}\;.
 \label{heffma}
\end{eqnarray}
Moreover, the stringMB can extract sub-blocks of matrices or their products corresponding to arbitrary active space. This feature is used to form matrix representations of the 
effective Hamiltonians $H^{\rm eff}$. 

\section{Results}

Owing to the memory requirements (associated with the storage of matrix representations of the operator) of the stringMB code, we can deal with relatively small systems yet epitomizing situations encountered in the calculations for larger systems and processes. For this reason, we employed ubiquitous models corresponding to the linear chains of hydrogen atoms: H6 and H8 models. 
By varying neighboring hydrogens distance ($R_{\rm H-H}$), one can smoothly transition from the single-reference character of the ground-state wave function for smaller $R_{\rm H-H}$ distances ($\simeq$ 2.0 a.u. or less) to quasi-degenerate regime (2.75 and 3.00 a.u.) - a  typical situation encountered in bond breaking/forming processes. 
These benchmark systems are commonly used in accuracy studies of ab-into methodologies to deal with the strong correlation effects. 
We used STO-3G basis set \cite{hehre1969self} and 
in all calculations, we used restricted Hartree-Fock (RHF) molecular bases composed of 6 and 8 molecular orbitals for H6 and H8 systems, respectively. 
For the strong correlation regime ($R_{\rm H-H}=3.0$ a.u.), there is no obvious choice of the active space for the ground-state problem; according to the wave function analysis for large separations, all orbitals fall into the category of "important" or "active" orbitals.
The results of our simulations for H6 and H8 systems are summarized in Tables \ref{tab1} - \ref{tab6}. 

In Table \ref{tab1}, we collected CC and DUCC results obtained for active space defined by two highest occupied orbitals (orbitals 2 and 3) and two lowest virtual orbitals (orbitals 4 and 5). The canonical CCSD, CCSDT, and CCSDTQ energies (corresponding columns are denoted as SD, SDT, and SDTQ) are compared to the FCI ones and lowest eigenvalues of the downfolded Hamiltonian in the active space for 
CCSD (DUCC-SD), CCSDT (DUCC-SDT), and CCSDTQ (DUCC-SDTQ) sources of the external amplitudes $T_{\rm ext}$ to calculate $\sigma_{\rm ext}(\mathfrak{h})$ according to formula (\ref{sigext}). While for weakly correlated regimes, all choices of
$T_{\rm ext}$ result in comparable accuracy with respect to the FCI energies, for larger $R_{\rm H-H}$ distances the quality of $T_{\rm ext}$ matters. For example, for larger distances, only $T_{\rm ext}$ defined at the CCSDTQ level yields DUCC-SDTQ energies of the CCSDTQ/FCI quality for all geometries. 
Other approaches, CCSD and CCSDT, for larger values of $R_{\rm H-H}$, provide energies significantly below the FCI ones. However, it is interesting to notice that the corresponding eigenvalues of DUCC-SD and DUCC-SDT can reinstate the variational character of ground-state energy despite using $T_{\rm ext}$ stemming from CCSD and CCSDT calculations, which is an important property of the DUCC approach.

\begin{table*}[!ht]
    \centering
    \begin{tabular}{l c c c c c c c}
    \hline \hline \\[-0.2cm]
   $R_{\rm H-H}$  &  FCI & SD & SDT & SDTQ & DUCC-SD & DUCC-SDT & DUCC-SDTQ  \\
        \hline \\[-0.1cm]
        1.50 & -3.199566 & -3.199332  & -3.199601  &  -3.199566 & -3.199324  & -3.199562  &  -3.199566   \\[0.1cm]
        1.75 & -3.245936 & -3.245603  & -3.246054  &  -3.245936 & -3.245547  & -3.245923  &  -3.245936   \\[0.1cm]
        2.00 & -3.217699 & -3.217277  & -3.218047  &  -3.217699 & -3.217040  & -3.217655  &  -3.217697   \\[0.1cm]
        2.25 & -3.156624 & -3.156266  & -3.157559  &  -3.156621 & -3.155447  & -3.156484  &  -3.156618   \\[0.1cm]
        2.50 & -3.085398 & -3.085691  & -3.087713  &  -3.085380 & -3.083217  & -3.084962  &  -3.085374   \\[0.1cm]
        2.75 & -3.016841 & -3.019512  & -3.022159  &  -3.016770 & -3.012642  & -3.015537  &  -3.016758   \\[0.1cm]
        3.00 & -2.957646 & -2.967326  & -2.969163  &  -2.957405 & -2.948732  & -2.953850  &  -2.957384    \\[0.1cm]
         \hline \hline
    \end{tabular}
      \caption{Comparison of energies of the downfolded Hamiltonians for the linear H6 system in the STO-3G basis set based on various sources of the external amplitudes $T_{\rm ext}$ used to approximate the $\sigma_{\rm ext}$ operator ($\sigma_{\rm ext}\simeq T_{\rm ext} - T_{\rm ext}^{\dagger}$). All simulations used restricted Hartree-Fock molecular orbitals 2,3, and 4,5 as active occupied and virtual orbitals, respectively. In the linear chain of the H atoms, the geometry is defined by the distance between neighboring hydrogen atoms ($R_{\rm H-H}$) in a.u.}
    \label{tab1}
\end{table*}
%
%
\begin{table*}[!ht]
    \centering
        \begin{tabular}{l c c c c c c c}
\hline \hline  \\[-.2cm]
        $R_{\rm H-H}$  &  FCI & SD & SDT & SDTQ & DUCC-SD & DUCC-SDT & DUCC-SDTQ  \\
        \hline \\[-0.1cm]
        1.50 & -4.235775 & -4.235111 & -4.235846 & -4.235775 & -4.235071 & -4.235757 & -4.235774 \\[0.1cm]
        1.75 & -4.315273 & -4.314347 & -4.315504 & -4.315273 & -4.314173 & -4.315222 & -4.315271 \\[0.1cm]
        2.00 & -4.286011 & -4.284844 & -4.286688 & -4.286013 & -4.284235 & -4.285862 & -4.286005 \\[0.1cm]
        2.25 & -4.208339 & -4.207232 & -4.210169 & -4.208337 & -4.205334 & -4.207876 & -4.208316 \\[0.1cm]
        2.50 & -4.114829 & -4.115000 & -4.119502 & -4.114795 & -4.109473 & -4.113350 & -4.114739 \\[0.1cm]
        2.75 & -4.023783 & -4.029321 & -4.035510 & -4.023578 & -4.013082 & -4.018712 & -4.023447 \\[0.1cm]
        3.00 & -3.944748 & -3.972672 & -3.978401 & -3.943920 & -3.912005 & -3.921323 & -3.943614 \\[0.1cm]
         \hline \hline
    \end{tabular}
      \caption{Comparison of energies of the downfolded Hamiltonians for the linear  H8 system in the STO-3G basis set based on various sources of the external amplitudes $T_{\rm ext}$ used to approximate the $\sigma_{\rm ext}$ operator ($\sigma_{\rm ext}\simeq T_{\rm ext} - T_{\rm ext}^{\dagger}$). All simulations used restricted Hartree-Fock molecular orbitals 2,3, and 4,5 as active occupied and virtual orbitals, respectively. In the linear chain of the H atoms, the geometry is defined by the distance between neighboring hydrogen atoms ($R_{\rm H-H}$) in a.u.}
    \label{tab2}
\end{table*}
The CC and DUCC results for various geometries of the H8 system are collected in Table \ref{tab2}. In this case, we also used active spaces defined by the two highest occupied orbitals (orbitals 3 and 4) and the two lowest virtual orbitals (orbitals 5 and 6). 
For the equilibrium geometry, including the triple and quadruple excitations in the external cluster operator, results in the near-FCI quality of the DUCC results. The accuracy of the DUCC-SD formalism is also satisfactory (less than 0.7 miliHartree error with respect to the FCI result), showing the effectiveness of DUCC formalism in compressing the dynamical correlation effects even in the case when a simple form of external amplitudes is invoked. 
In analogy to the H6 case, one can observe the variational breakdown of the CCSD and CCSDT results for the larger H-H separations and restoration of the variational character by the corresponding DUCC formulations. 

The DUCC case of downfolding is active-space specific (the so-called cluster amplitudes universal problem discussed in Ref.\cite{kowalski2021dimensionality}). However, to explore to what extent DUCC downfolding formalism depends on the active space definition, in Table \ref{tab3}, we collected DUCC-CCSDTQ energies  for the H8 model with $R_{\rm H-H}$=2.0 a.u. and three model spaces defined  by 
$\lbrace 3,4,5,6 \rbrace$, $\lbrace 2,3,6,7 \rbrace$, and $\lbrace 1,2,7,8 \rbrace$ orbitals. One can see that although the DUCC-CCSDTQ energies are not equal, the energy discrepancies between results corresponding to various active spaces do not exceed 0.016 milliHartree. 

\begin{table*}[!ht]
    \centering
    \begin{tabular}{l c c c }
\hline\hline \\
        FCI  &   DUCC-CCSDTQ &
                 DUCC-CCSDTQ & 
                 DUCC-CCSDTQ \\
             &   ($R=\lbrace 3,4 \rbrace$,$S=\lbrace 5,6 \rbrace$) 
             &   ($R=\lbrace 2,3 \rbrace$,$S=\lbrace 6,7 \rbrace$)
             &   ($R=\lbrace 1,2 \rbrace$,$S=\lbrace 7,8 \rbrace$) \\[0.1cm]
        \hline \\
       -4.286011 & -4.286005 & -4.285865 & -4.285853 \\[0.1cm]
         \hline\hline
    \end{tabular}
    \caption{The DUCC-CCSDTQ results for the H8 model ($R_{\rm H-H}$=2.0 a.u.) were obtained with the STO-3G basis set for various choices of active spaces.}
    \label{tab3}
\end{table*}

The effect of the size of the active space is discussed in Table  \ref{tab4} where we collated DUCC-CCSDTQ results obtained for various geometries of H8 and large $\lbrace 2,3,4,5,6,7 \rbrace$-generated active space. As expected, the increase in the active space size results in more accurate DUCC-CCSDT energies, especially for large distances. It is also evident that the DUCC-CCSDTQ results surpass the CCSDTQ accuracy for $R_{\rm H-H}$=3.0 a.u.  due to the fact that higher-than-quadruple excitations are included in the diagonalization of the downfolded Hamiltonian in the active space. This behavior reflects 
the fact discussed in Ref.\cite{bauman2022coupled2c} that size of the active space can compensate for the missing correlation effects not included in the external cluster operators. 

\begin{table*}[!ht]
    \centering
    \begin{tabular}{l c c c}
\hline\hline \\
    $R_{\rm H-H}$   &  FCI &  CCSDTQ & DUCC-CCSDTQ  \\
        \hline \\
        2.00 & -4.286011 & -4.286013  & -4.286008 \\[0.1cm]
        2.50 & -4.114829 & -4.114795  & -4.114782 \\[0.1cm]
        3.00 & -3.944748 & -3.943920  & -3.944137 \\[0.1cm]
         \hline\hline
    \end{tabular}
   \caption{The DUCC-CCSDTQ results were obtained for various geometries of the H8 model in the STO-3G basis set for $\lbrace 2,3,4,5,6,7 \rbrace$-generated active space (see text for details).}
    \label{tab4}
\end{table*}

From the point of view of quantum computing application, it is essential to evaluate the performance of techniques to approximate the many-body forms of the downfolded Hamiltonians and the potential accuracies of quantum solvers. To this end, we will analyze the finite rank expansion for 
Eq.(\ref{comm1}) and variant of the Connected Methods Expansion (CMX) \cite{horn1984t,cioslowski1987connected,knowles1987validity,mancini1994analytic,noga2002use}
based on the Peeters-Devreese-Soldatov (PDS) functional 
\cite{peeters1984upper,soldatov1995generalized}, which has been recently explored in the context of quantum computing.\cite{kowalski2020quantum,peng2021variational,claudino2021improving}
In this paper we will apply low-rank PDS formulations
PDS(3) and  PDS(4)
(for details we refer the reader to Refs.\cite{peeters1984upper,soldatov1995generalized}) to identify ground-state energies of the downfolded Hamiltonians. 
It is worth noting that the low ranks of the PDS approach 
can be effectively implemented on quantum computers. The PDS results are collated in Table \ref{tab6}, providing Hartree-Fock, complete active space self-consistent field (using four active electrons distributed over four active orbitals), and active-space FCI  energies. For H6 and H8 models, we used 
$\lbrace 2,3,4,5 \rbrace$- and 
$\lbrace 4,5,6,7 \rbrace$-generated active spaces, respectively. 
Before discussing the PDS results, we should stress the efficiency of the downfolding procedures (here illustrated in the example of the DUCC-CCSDTQ approach) in capturing the out-of-active-space correlation effect. This is best illustrated by comparing DUCC-CCSDTQ vs. CASSCF(4,4) and Active-space FCI energies. Despite using the same active space definitions, the CASSCF(4,4) and Active-space FCI energies for all geometries of H6 and H8, in contrast to the DUCC-CCSDTQ approach, are characterized by significant errors with respect to the FCI energies. Although, in many cases, quantum simulations are performed for small dimensionality active spaces using bare Hamiltonians, the quality of the results can be significantly improved without a significant increase in quantum computing resources by using a downfolded form of the Hamiltonian. 
As seen from Table \ref{tab5}, the PDS(3) can provide much better quality results than active-space FCI. The PDS(4) can further refine the accuracies of the PDS(3) approach reducing the errors with respect to exact DUCC-CCSDTQ energies to within 0.7 milliHartree for H6 and H8 model systems. 
\begin{table*}[!ht]
    \centering
    \begin{tabular}{l c c c c }
    \hline \hline \\
  Method & H6  & H6  & H8 &
  H8   \\
  &  ($R_{\rm H-H} = 2.0$ a.u) & ($R_{\rm H-H} = 3.0$ a.u) & ($R_{\rm H-H} = 2.0$ a.u) &
   ($R_{\rm H-H} = 3.0$ a.u) \\
  \hline \\
   HF        & -3.105850 & -2.675432 & -4.138199 & -3.572347    \\[0.1cm]
   CASSCF(4,4) & -3.175370  & -2.856832 & -4.205528 & -3.699677 \\[0.1cm]
   Active-space FCI & -3.166938    &  -2.802092    &   -4.190602   &  -3.665605    \\[0.1cm]
   FCI       & -3.217699 & -2.957646 & -4.286011 &  -3.944748   \\[0.1cm]
   DUCC-CCSDTQ & -3.217697  & -2.957384  & -4.286005 &  -3.943614 \\[0.1cm]
   PDS(3)    & -3.214888 & -2.953067 & -4.283332 &  -3.941223   \\[0.1cm]
   PDS(4)    & -3.217234 & -2.956712 & -4.285622 &  -3.943349   \\[0.1cm]   
         \hline \hline
    \end{tabular}
           \caption{Comparison of the  CASSCF(4,4), active-space FCI, DUCC-CCSDTQ, and PDS(3)/PDS(4) energies for H6 and H8 model systems. The PDS(3)/PDS(4) approaches were applied to evaluate the ground-state energy of the DUCC-CCSDTQ effective Hamiltonians. The $\lbrace 2,3,4,5 \rbrace$- and 
$\lbrace 4,5,6,7 \rbrace$-generated active spaces were used for H6 and H8 systems, respectively.}
    \label{tab5}
\end{table*}
In the last part of our discussion, we analyze the accuracies of the finite rank (${\rm Max}_R$) approximations for downfolded Hamiltonians. The results for H6 and H8 models are collected in Table \ref{tab6}. The convergence of the commutator expansions is illustrated in the example of the 10th-rank commutator expansion. In all cases discussed in Table \ref{tab6} 
${\rm Max}_R$=10 approximation reproduces virtually the exact DUCC-CCSDTQ energies. In practical applications based on the many-body form of the downfolded Hamiltonian, only low-rank commutator expansions (${\rm Max}_R$=1,...,4) are numerically feasible (the ${\rm Max}_R$=0 corresponds to the active-space FCI results). One can observe that for weakly correlated situations ($R_{\rm H-H}$=2.0 a.u.) ${\rm Max}_R$=3 provides a satisfactory approximation of the exact DUCC-CCSDTQ energies. For strongly correlated case ($R_{\rm H-H}$=3.0 a.u.), the inclusion of the 4-th rank commutators (${\rm Max}_R$=4) is needed. In practical application, however, all expansions are based on the mixing of ${\rm Max}_R=n$ contributions with 
$n+1$-rank commutators stemming from the Fock matrix terms to reinstate the so-called perturbative balance (see Refs.\cite{} for the discussion). For example, ${\rm Max}_R$=1 case epitomizes a situation where the perturbative balance is violated, and non-variational energies can be obtained. Therefore for the strongly correlated cases, we recommend the expansions based on the inclusion of ${\rm Max}_R$=3 or ${\rm Max}_R$=4 terms with the Fock-operator-dependent terms originating in the 4-th and 5-th commutators, respectively. 
\begin{table*}[!ht]
    \centering
    \begin{tabular}{l c c c c }
    \hline \hline \\
  ${\rm Max_R}$/Method & H6  & H6  & H8 &
  H8   \\
  &  ($R_{\rm H-H} = 2.0$ a.u) & ($R_{\rm H-H} = 3.0$ a.u) & ($R_{\rm H-H} = 2.0$ a.u) &
   ($R_{\rm H-H} = 3.0$ a.u) \\
  \hline \\
   ${\rm Max_R}$=0   & -3.166938 & -2.802092 & -4.190602 &  -3.665605  \\[0.1cm]
   ${\rm Max_R}$=1  & -3.269110 & -3.116145 & -4.382423 &  -4.228605  \\[0.1cm]
   ${\rm Max_R}$=2  & -3.218732 & -2.976207 & -4.288761 &  -3.986313  \\[0.1cm]
   ${\rm Max_R}$=3  & -3.217344 & -2.949796 & -4.285044 &  -3.927241  \\[0.1cm]
   ${\rm Max_R}$=4  & -3.217693 & -2.956814 & -4.285985 &  -3.941744  \\[0.1cm]
   ${\rm Max_R}$=5  & -3.217699 & -2.957632 &  -4.286012 & -3.944383   \\[0.1cm]
   ${\rm Max_R}$=10 &  3.217697 &  -2.957384  & -4.286005  &  -3.943615  \\[0.1cm]
   \hline \\[-0.2cm]   
   DUCC-CCSDTQ &-3.217697  & -2.957384  & -4.286005 &  -3.943614 \\[0.1cm]
FCI       & -3.217699 & -2.957646 & -4.286011 &  -3.944748   \\[0.1cm]
         \hline \hline
    \end{tabular}
    \caption{Comparison of energies obtained with finite Commutator Expansion (CCSDTQ based) for the downfolded Hamiltonians and   Exact Downfolding(DUCC-CCSDTQ)  for H6 and H8 models in the STO-3G basis set.}
    \label{tab6}
\end{table*}
%
%
%

\section{Conclusion}

A series of calculations were conducted to examine the impact of approximations made on external cluster amplitudes on CC downfolded energies. Simple model systems, H6 and H8 linear chains, were utilized to continuously vary the extent of correlation effects from weakly to strongly correlated regimes (with $R_{\rm H-H}$ ranging from 2.0 to 3.0 a.u.). The results showed that while the external cluster amplitudes from SR-CCSD calculations were satisfactory for the weakly correlated situation, for the strongly correlated case, the effect of triply and quadruply excited external clusters could no longer be neglected. The downfolding procedure acted as stabilizers and could restore the variational character of energies despite the fact that external amplitudes were obtained from SR-CC calculations that suffer from the variational collapse. Furthermore, the downfolded energies obtained for various active spaces on the H8 systems (including those that did not include essential correlation effects) had only small discrepancies, demonstrating the approximate invariance of downfolding energies for various active spaces. This was the case for all SES-CC types active spaces for commutative SR-CC formulations. Additionally, it was shown that the downfolded Hamiltonians could be effectively diagonalized with low-order PDS formulations for both weakly and strongly correlated regimes.

The assessment of the impact of the maximum rank of the commutator expansion on the precision of downfolded energies is an integral aspect of the analysis associated with practical applications of CC downfolding procedures. Our findings demonstrate that an increase in the degree of correlation effects necessitates the inclusion of higher rank commutators. In particular, for H6/H8 $R_{\rm H-H}=2.0$ a.u., the inclusion of single, double, and triple commutators is sufficient to achieve a favorable agreement with the FCI energies. However, for $R_{\rm H-H}=3.0$ a.u., higher-order commutators (quadruple/pentuple) must be incorporated into the approximation. It should be noted, however, that for the $R_{\rm H-H}=3.0$ a.u. scenario, an active space that can distinguish between static and dynamical correlation effects cannot be constructed (i.e. all orbitals must be considered active). Nonetheless, as Table (\ref{tab6}) demonstrates, the corresponding commutator-rank expansion rapidly converges to the FCI energies.

\section{Acknowledgement}

This  work was  supported by the Quantum Science Center (QSC), a National Quantum Information Science Research Center of the U.S. Department of Energy (under FWP  76213) and by 
``Embedding QC into Many-body Frameworks for Strongly Correlated Molecular and Materials Systems''  project, which is funded by the U.S. Department of Energy, Office of Science, Office of Basic Energy Sciences, the Division of Chemical Sciences, Geosciences, and Biosciences (under FWP 72689). This work used resources from the Pacific Northwest National Laboratory (PNNL).
PNNL is operated by Battelle for the U.S. Department of Energy under Contract DE-AC05-76RL01830.

\section*{AUTHOR DECLARATIONS}
\subsection*{Conflict of Interest}
The authors have no conflicts of interest to declare.

\section*{DATA AVAILABILITY}
The data that support the findings of this study are available from the corresponding authors
upon reasonable request.

\clearpage

\bibliography{ref3.bib}

\end{document}